\documentclass[prc,twocolumn,10pt]{revtex4}

\usepackage{draftcopy}
\usepackage{epsfig}
\newcommand{\Version}       {Version: High-pt42.tex, July 1, 2003}
\newcommand{\bnl}           {$\rm^{1}$}
\newcommand{\ires}          {$\rm^{2}$}
\newcommand{\kraknuc}       {$\rm^{3}$}
\newcommand{\krakow}        {$\rm^{4}$}
\newcommand{\baltimore}     {$\rm^{5}$}
\newcommand{\newyork}       {$\rm^{6}$}
\newcommand{\nbi}           {$\rm^{7}$}
\newcommand{\texas}         {$\rm^{8}$}
\newcommand{\bergen}        {$\rm^{9}$}
\newcommand{\bucharest}     {$\rm^{10}$}
\newcommand{\kansas}        {$\rm^{11}$}
\newcommand{\oslo}          {$\rm^{12}$}

\begin{document}

\title{Transverse Momentum Spectra in Au+Au and d+Au Collisions at
  $\sqrt{s_{NN}}$=200 GeV and the Pseudorapidity Dependence of High p$_T$ Suppression}

\date{\Version}

\author{
  I.~Arsene\bucharest,
  I.~G.~Bearden\nbi, 
  D.~Beavis\bnl, 
  C.~Besliu\bucharest, 
  B.~Budick\newyork, 
  H.~B{\o}ggild\nbi, 
  C.~Chasman\bnl, 
  C.~H.~Christensen\nbi, 
  P.~Christiansen\nbi, 
  J.~Cibor\kraknuc, 
  R.~Debbe\bnl, 
  E. Enger\oslo,  
  J.~J.~Gaardh{\o}je\nbi, 
  M.~Germinario\nbi, 
  K.~Hagel\texas, 
  O.~Hansen\nbi, 
  A.~Holm\nbi, 
  H.~Ito\bnl$^,$\kansas, 
  A.~Jipa\bucharest, 
  F.~Jundt\ires, 
  J.~I.~J{\o}rdre\bergen, 
  C.~E.~J{\o}rgensen\nbi, 
  R.~Karabowicz\krakow, 
  E.~J.~Kim\bnl, 
  T.~Kozik\krakow, 
  T.~M.~Larsen\oslo, 
  J.~H.~Lee\bnl, 
  Y.~K.~Lee\baltimore, 
  S.~Lindal\oslo, 
  G.~Lystad\bergen, 
  G.~L{\o}vh{\o}iden\oslo, 
  Z.~Majka\krakow, 
  A.~Makeev\texas, 
  B.~McBreen\bnl, 
  M.~Mikelsen\oslo, 
  M.~Murray\texas$^,$\kansas, 
  J.~Natowitz\texas, 
  B.~Neumann\kansas, 
  B.~S.~Nielsen\nbi, 
  J.~Norris\kansas, 
  D.~Ouerdane\nbi, 
  R.~P\l aneta\krakow, 
  F.~Rami\ires, 
  C.~Ristea\bucharest, 
  O.~Ristea\bucharest, 
  D.~R{\"o}hrich\bergen, 
  B.~H.~Samset\oslo, 
  D.~Sandberg\nbi, 
  S.~J.~Sanders\kansas, 
  R.~A.~Scheetz\bnl, 
  P.~Staszel\nbi$^,$\krakow, 
  T.~S.~Tveter\oslo, 
  F.~Videb{\ae}k\bnl, 
  R.~Wada\texas, 
  Z.~Yin\bergen, 
  I.~S.~Zgura\bucharest\\ 
  The BRAHMS Collaboration \\ [1ex]
  \bnl~Brookhaven National Laboratory, Upton, New York 11973, USA\\
  \ires~Institut de Recherches Subatomiques and Universit{\'e} Louis
  Pasteur, Strasbourg, France\\
  \kraknuc~Institute of Nuclear Physics, Krakow, Poland\\
  \krakow~M. Smoluchkowski Inst. of Physics, Jagiellonian University,
  Krakow, Poland\\
  \baltimore~Johns Hopkins University, Baltimore 21218, USA\\
  \newyork~New York University, New York 10003, USA\\
  \nbi~Niels Bohr Institute, Blegdamsvej 17, University of Copenhagen,
  Copenhagen 2100, Denmark\\
  \texas~Texas A$\&$M University, College Station, Texas, 17843, USA\\
  \bergen~University of Bergen, Department of Physics, Bergen,
  Norway\\
  \bucharest~University of Bucharest, Romania\\
  \kansas~University of Kansas, Lawrence, Kansas 66045, USA \\
  \oslo~University of Oslo, Department of Physics, Oslo, Norway\\
 }

\begin{abstract}
  We present spectra of charged hadrons from Au+Au and d+Au collisions
  at $\sqrt{s_{NN}}=200$ GeV measured with the BRAHMS experiment at
  RHIC.  The spectra for different collision centralities are compared
  to spectra from ${\rm p}+\bar{{\rm p}}$ collisions at the same
  energy scaled by the number of binary collisions.  The resulting
  ratios (nuclear modification factors) for central Au+Au collisions
  at $\eta=0$ and $\eta=2.2$ evidence a strong suppression in the high
  $p_{T}$ region ($>$2 GeV/c).  In contrast, the d+Au nuclear
  modification factor (at $\eta=0$) exhibits an enhancement of the
  high $p_T$ yields.  These measurements indicate a high energy loss
  of the high $p_T$ particles in the medium created in the central
  Au+Au collisions. The lack of suppression in d+Au collisions makes
  it unlikely that initial state effects can explain the suppression
  in the central Au+Au collisions.

  PACS numbers: 25.75 Dw.
\end{abstract}

\maketitle

Collisions between heavy nuclei in the energy domain now accessible at
the Relativistic Heavy Ion Collider (RHIC) are expected to lead to the
formation of an extremely hot high-density region possibly exhibiting
features characteristic of quark deconfinement, {\em i.e.} the
quark-gluon plasma (QGP). The first experiments with Au+Au collisions
at $\sqrt{s_{NN}}=200$ GeV suggest that very high energy densities
($\epsilon > 5$ GeV/fm$^3$) are achieved in the initial stages of
such collisions.  Furthermore, the reaction mechanism at RHIC has new
features as compared to lower energies, indicating a high degree of
nuclear transparency (as may be deduced from the low net proton
rapidity density measured in the region around midrapidity ($|y|<2$)
\cite{BRAHMSnetproton}). This resembles the scenario proposed by
Bjorken~\cite{Bjorken83}, in which the colliding nuclei suffer a
moderate relative rapidity loss and where subsequent particle
production in the boost invariant midrapidity region arises primarily
from quark-antiquark pair production from the breaking of color
strings between the interacting partons. Additionally, studies of the
particle production~\cite{BRAHMSmult,PHOBOSmult200} and the dynamics
of the expanding hadronic cloud that subsequently forms suggest that
the system, at least in later stages of the collision, may be in
thermal and chemical equilibrium. An analysis of particle ratios at
midrapidity indicates that the baryochemical potential is low
($\mu_{B} < 30$ MeV)~\cite{BRAHMSratio200}. This set of
observations naturally leads to speculation about whether a high
density deconfined state of quarks and gluons is indeed formed in
Au+Au collisions at RHIC.

In order to investigate the conditions prevailing early in the
evolution of the system it has been proposed
~\cite{Bjorken83,Gyulassy90,Baier95} that high momentum particles may
be a good probe of the conditions of the produced medium. Such
particles are associated with jet production from initial hard
parton scatterings and are predicted to suffer energy loss due to induced
gluon radiation as they traverse a medium with a high density of color
charges, resulting in a depletion of the high transverse momentum
component of their spectra. This high $p_T$ suppression has been of
much recent interest
~\cite{BRAHMS-qm2002CEJ,STARptAuAu,PHENIXptAuAu,PHOBOSptAuAu}.

In this Letter, we report on measurements of charged hadrons from
Au+Au collisions at $\sqrt{s_{NN}}=200$ GeV at pseudorapidities $\eta
= -ln(tan(\theta/2))= 0$ and $\eta=2.2$, where $\theta$ is the angle
of emission relative to the beam direction. The spectra, which have
been measured as a function of the collision centrality, are compared
to reference data from elementary ${\rm p}+\bar{{\rm p}}$ collisions
at the same energy using a scaling to our acceptance and to the
estimated number of binary collisions. We have also measured similar
spectra (for minimum bias (MB) collisions) for the reaction d+Au at
$\sqrt{s_{NN}}=200$ GeV in order to probe the possible role of initial
state effects and the influence of the participant volume.  For
central ($0-10\%$) Au+Au collisions we find a strong suppression of
the high transverse momentum component ($p_T > 2$ GeV/c) of the
spectra as compared to the scaled ${\rm p}+\bar{{\rm p}}$
spectra. This suppression diminishes significantly as the collision
centrality decreases. In contrast, an enhancement is observed for the
d+Au collisions.

\begin{figure}[htp]
  \epsfig{file=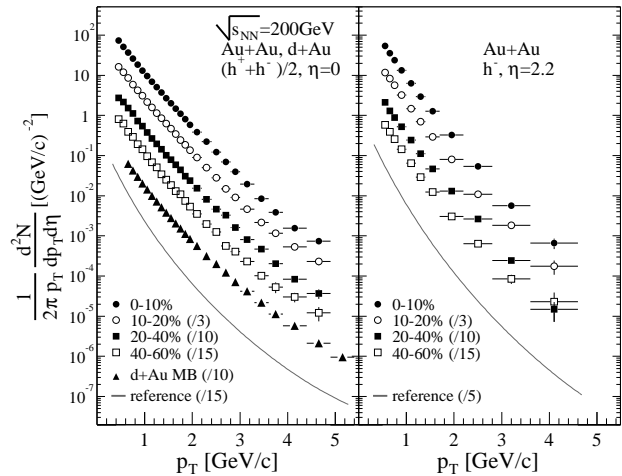,width=\linewidth}
  \caption{Invariant spectra of charged hadrons from Au+Au collisions
    at $\sqrt {s_{NN}} = 200$ GeV for pseudorapidities $\eta=0$ (left
    panel) and $\eta=2.2$ (right panel). Various centrality cuts are
    shown for Au+Au. The ${\rm p}+\bar{{\rm p}}$ reference spectra (appropriately acceptance 
    scaled) are shown for comparison.  The left panel also shows the
    d+Au spectrum.  For clarity, some spectra have been divided by the
    indicated factors.}
 \label{fig1}
 \vspace{-2mm}
\end{figure}

BRAHMS consists of two magnetic spectrometers (the MidRapidity
Spectrometer, MRS, and the Forward Spectrometer, FS) that for the
present measurements were positioned at $90^\circ$ (MRS) and $12^\circ$ (FS)
relative to the beam direction and measured hadrons ($h^+$) and
antihadrons ($h^-$) at pseudorapidities in the ranges $|\eta|< 0.1$ and
$2.1 < \eta < 2.3$, respectively. In addition, a set of global detectors were
used for miminum bias trigger and event characterization. In the
Au+Au run this trigger selected approximately 95\% of the Au+Au
interaction cross section. An additional hardware trigger selected
the $\approx25\%$ most central events for parts of the Au+Au
run. In the d+Au run, scintillator counters (INEL) were placed
around the nominal intersection point (IP) at $z= \pm1.6, \pm4.2$
and $\pm 6.6{\rm{m}}$, and used as the minimum bias trigger,
selecting $\approx 91\% \pm3\%$ of the $2.4 b$ d+Au inelastic cross
section. Spectrometer triggers were added to enhance the track
sample. Further details of the experimental setup and operation
can be found in refs.~\cite{BRAHMSNIM,BRAHMSmult}. Centrality
selection for the Au+Au collisions was done using multiplicity
detectors positioned around the IP. The IP position is determined
with a precision of $\sigma<0.9$ cm by the use of arrays of beam
counters (BB) placed at $z=\pm 2.2$m. For the d+Au reaction
study the vertex measurement by the INEL counters has a resolution
of approximately 9 cm.

Figure 1 shows measured invariant spectra for charged hadrons $(h^+ +
h^-)/2$ at $90^\circ$ (left panel) and for negative hadrons $(h^-)$ at
$12^\circ$ (right panel), corresponding to $\eta=0$ and $2.2$. The
displayed spectra for Au+Au collisions are for centralities of
0--10\%, 10--20\%, 20--40\%, and 40--60\%. The spectra are from
measurements at various magnetic fields and have been corrected for
the acceptance of the spectrometers and for the tracking
efficiency. No decay corrections have been applied. Also shown in the
figure is our measured spectrum from d+Au collisions. The reference is
from ${\rm p}+\bar{{\rm p}}$ collisions measured in the range
$|\eta|<2.5$ by the UA1 experiment at CERN~\cite{UA1}. To
compare with our spectra we apply a $p_T$ and $\eta$ dependent
correction estimated using the HIJING (v.\ 1.383)
code~\cite{hijing}, which reproduces the main features of ${\rm
p}+{\rm p}$ collisions well. We have compared our corrected spectrum
at $\eta=0$ with the $\sqrt{s}$=200 GeV ${\rm p}+{\rm p}$ distribution
recently measured by the STAR collaboration ~\cite{Starpp} and find
excellent agreement. No similar comparison is available for the more
forward rapidity.  Consequently, we use the model--scaled ${\rm
p}+\bar{{\rm p}}$ spectrum for comparison, noting that HIJING predicts
a $p_T$ dependent difference between $h^-$ and $(h^+ + h^-)/2$ at 
$\eta=2.2$. This has been taken into account in constructing the $h^-$
reference spectrum used for the $\eta=2.2$ analysis.

A useful way to compare the spectra from nucleus-nucleus
collisions to those from nucleon-nucleon collisions is to scale the
normalized ${\rm p}+{\rm p}$ spectrum (assuming
$\sigma_{inel}^{pp}=42mb$) by the number of binary collisions
($N_{bin}$) corresponding to the centrality cuts applied to the
nucleus-nucleus spectra and construct the ratio. This ratio is called
the nuclear modification factor,
$R_{AA}=(\sigma_{inel}^{pp}/N_{bin})(d^2N^{AA}/dp_Td\eta)/(d^2\sigma^{pp}/dp_Td\eta)$.

\begin{figure}[htp]
\vspace{2mm}
  \epsfig{file=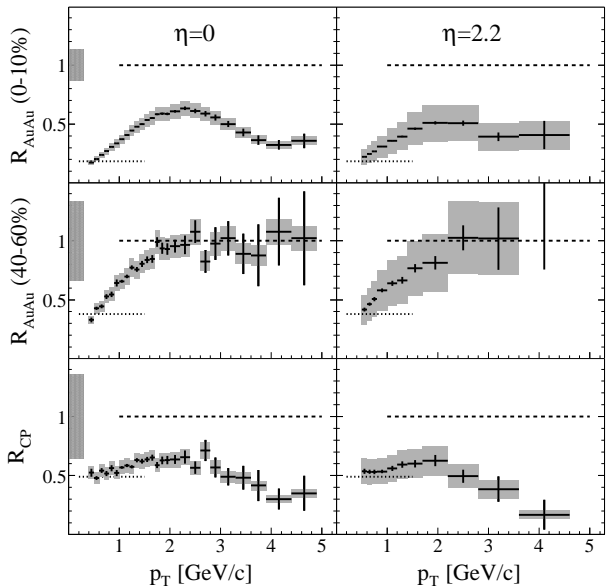,width=\linewidth}
  \caption{ Top row: Nuclear modification factors $R_{AuAu}$ as a
    function of transverse momentum for Au+Au collisions at $\eta=0$
    and $\eta=2.2$ for the $0-10\%$ most central collisions. Middle
    row: as top row, but for centralities $40-60\%$. Bottom row: ratio
    of the $R_{AuAu}$ factors for the most central and most peripheral
    collisions at the two rapidities. The dotted and dashed lines show
    the expected value of $R_{AuAu}$ using a scaling by the number of
    participants and by the number of binary collisions,
    respectively. Error bars are statistical.  The grey bands indicate
    the estimated systematic errors.  The grey band at $p_T=0$ is the
    uncertainty on the scale.}
  \label{fig2}
\end{figure}

Figure 2 (upper two rows) shows the ratios $R_{AuAu}$, as a function
of $p_T$ for different centrality cuts for the Au+Au measurements at
$\eta=0$ and $2.2$. For the most central ($0-10\%$) bin we use
$N_{bin}= 897 \pm 117$, and for the most peripheral ($40-60\%$)
$N_{bin}=78\pm 26$. For the d+Au reaction we have used $N_{bin}=7.2\pm
0.3$.  The $R_{AuAu}$ rise from values of 0.2--0.4 at low $p_T$ to a
maximum at $p_T\approx 2$ GeV/c. The low $p_T$ part of the spectrum is
associated with soft collisions and should therefore scale with the
number of participants. Thus the applied scaling with the (larger)
$N_{bin}$ value reduces $R_{AuAu}$ at the lower $p_T$. Beyond $p_T
\approx 2$ GeV/c, $R_{AuAu}$ is expected to be close to 1. In fact,
measurements at CERN-SPS for $\sqrt {s_{NN}} = 17 $ GeV collisions for
neutral pions~\cite{WA98-RAA}, negative hadrons~\cite{NA49-RAA} and
charged pions~\cite{CERES-RAA} show that $R_{AA}$ is equal to 1 at
$p_T=1.5$ GeV/c and increases to about 1.5 at $p_T=3$
GeV/c~\cite{WANG01}. This is the Cronin effect~\cite{Cronin75}
attributed to multiple scattering of partons in the initial stages of
the collision. Above $p_T\approx2$ GeV/c the $R_{AuAu}$ distributions
shown in fig.\ 2 decrease and are systematically lower than unity for
the central collisions, while they remain near 1 for more peripheral
collisions.  Indeed, for the most central collisions at both
pseudorapidities, $R_{AuAu}$ is only about 0.4 at $p_T \approx 4$
GeV/c. The high $p_T$ component of the Au+Au spectra is therefore
suppressed by a factor of 3-4 as compared to the SPS results at lower
energies and by a factor of two compared to simple binary scaling. We
note, however, that because we lack an independent measurement of the
${\rm p}+{\rm p}$ reference spectrum at forward rapidity, the
systematic error on $R_{AuAu}$ at $\eta=2.2$ is estimated to be
$\sigma_{sys}\approx 30\%$ at high $p_T$. In order to remove this
large, and model dependent systematic error from our reference
spectra, we form the ratio $R_{cp} =
\frac{N_{bin}(P)}{N_{bin}(C)}\times[\frac{d^2N}{p_Tdp_Td\eta}|_C]/[\frac{d^2N}{p_Tdp_Td\eta}|_P]$,
where 'C' and 'P', denote the most central and peripheral bins,
respectively. As can be seen in the bottom panels of fig.\ 2, $R_{cp}$
shows a clear decrease for $p_T > 2 $~GeV/c for both $\eta=0$ and
$\eta=2.2$.

\begin{figure}[htp]
  \epsfig{file=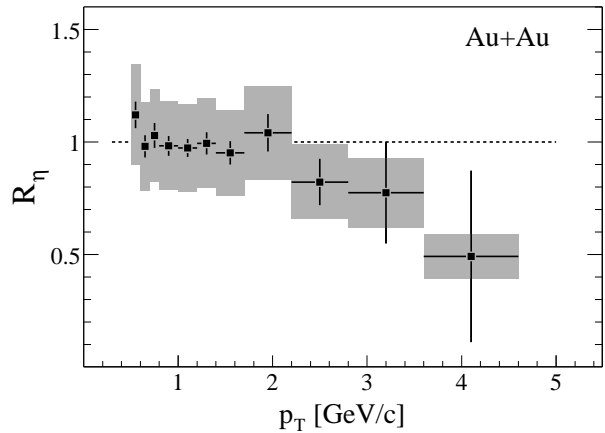,width=\linewidth}
  \caption{Ratio $R_\eta$ of $R_{cp}$ distributions at $\eta=2.2$ and
  $\eta=0$. Statistical errors are indicated by the bars, while systematic errors are shown by the grey bands.} 
  \label{fig3}
\end{figure}

We investigate the pseudorapidity dependence of the high $p_T$
suppression using the ratio $R_\eta=
R_{cp}(\eta=2.2)/R_{cp}(\eta=0)$. This ratio, shown in fig.\ 3, is
free from systematic errors arising from the determination of the
$N_{bin}$ values corresponding to the centrality cuts. Figure 3 shows
that, within errors, the degree of high $p_{T}$ suppression (for $p_T
> 2$ GeV/c) observed at $\eta=2.2$ is similar to or larger than at
$\eta=0$. We note that one might naively expect the suppression to be
proportional to the measured $dN/d\eta$ for charged particles which in
turn is expected to be proportional to $dN_{gluon}/d\eta$.  Since the
measured $dN/d\eta$ distributions~\cite{BRAHMSmult, PHOBOSmult200} are
roughly flat in this rapidity region, one would expect similar
suppression factors. However, the strength of the underlying
suppression mechanism (partonic or hadronic) may depend on the
pseudorapidity density, wherefore it is important to explore the
details of $p_T$ suppression over a wide rapidity region.

For comparison to the results obtained for the Au+Au collisions we
have investigated the d+Au reaction at the same energy at
$\eta=0$. In fig.\ 4 we present the corresponding $R_{dAu}$
distribution, analyzed in the same way as the Au+Au
collisions. We have applied no centrality cuts, so the
distribution reflects our minimum bias collision data.
It is striking that $R_{dAu}$ shows no suppression of
the high $p_T$ component. Rather, it shows an
enhancement, similar to the one observed at lower energies.

\begin{figure}[htp]
  \epsfig{file=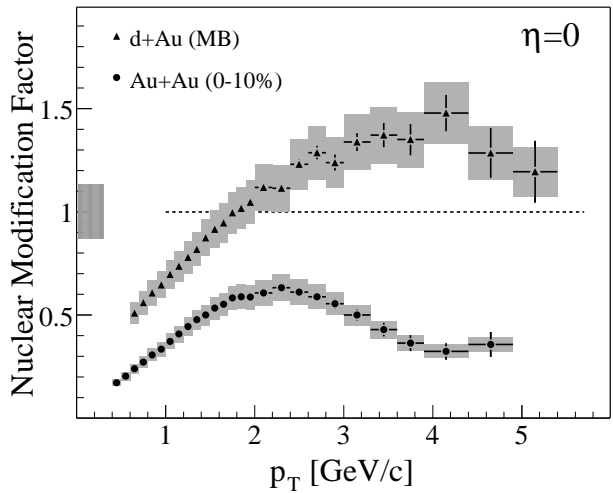,width=\linewidth}
 \vspace{-5mm}
  \caption{Nuclear modification factor measured for minimum bias
    collisions of d+Au at $\sqrt {s_{NN}} = 200$ GeV compared to
    central Au+Au collisions.  Error bars represent statistical
    errors. Systematic errors are denoted by the shaded bands.  The
    grey band at $p_T=0$ is the uncertainty on the scale of
    $R_{AuAu}$.}
  \label{fig4}
\end{figure}

From these measurements we conclude that central collisions between
Au+Au nuclei exhibit a very significant suppression of the high
transverse momentum component as compared to nucleon--nucleon
collisions. This suppression appears to be directly correlated with
the size of the participant volume, as demonstrated by the fact that
the much smaller participant volume resulting from the d+Au collisions
shows no suppression and by the fact that more peripheral collisions
between the Au nuclei show less suppression than the corresponding
central collisions. It is reasonable to surmise that the effect is
related to medium effects tied to a large volume with high energy
density.  It has been proposed~\cite{gluonsat} that gluon saturation
effects in the colliding Au+Au nuclei, {\it i.e.} initial state effects
resulting from the high laboratory energy of the colliding nuclei
might limit the phase space available for the production of high
momentum particles. Such an explanation appears improbable in view of
the results for the d+Au measurements which utilize projectiles at the
same energy.

In summary, the BRAHMS measurements demonstrate a significant
suppression of the high $p_T$ component of transverse momentum spectra
for hadrons measured at two rapidities for Au+Au collisions at
$\sqrt{s_{NN}}=200$~GeV. The suppression is seen to diminish with
decreasing collision centrality and is absent in d+Au collisions. In
fact, the observation of a Cronin enhancement in d+Au reaction seems
to exclude possible initial state effects contributing to the observed
suppression in collisions between large nuclei. We conclude that the
observed suppression in Au+Au is consistent with significant medium
effects in the most violent collisions, {\em i.e.} those that have the
largest participant volumes.   The persistance of the suppression
to $\eta=2.2$ suggests that the volume which causes the suppression is
extended also in the longitudinal direction. Whether the observed
suppression is tied to absorption or energy loss of scattered high
momentum partons by a dense partonic medium ~\cite{Vitev03}, to
absorption at a later hadronic stage~\cite{Gallmeister03} or to some
other mechanism is as yet unclear. This clearly warrants further
systematic investigations, notably by studying the effect over the
largest possible rapidity range with identified particles in order to
probe varying source conditions and absorption mechanisms and by
carrying out experiments at lower beam energies.

This work was supported by the division of Nuclear Physics of the
Office of Science of the U.S. DOE, the Danish Natural Science
Research Council, the Research Council of Norway, the Polish State
Com. for Scientific Research and the Romanian Ministry of
Research.

\newpage
\end{document}